\documentclass[10pt, conference]{IEEEtran}
\ifCLASSINFOpdf
  \usepackage[pdftex]{graphicx}
  \graphicspath{{../pdf/}{../jpeg/}}
  \DeclareGraphicsExtensions{.pdf,.jpeg,.png}
\else
\fi
\usepackage{fixltx2e}

\usepackage{stfloats}
\usepackage{mathtools} 
\usepackage{tabulary}
\usepackage{booktabs}

\usepackage{amsmath}
\usepackage{amsthm}
 \usepackage{url} 
\usepackage{float}
\restylefloat{table}
\usepackage{subfig}
\usepackage{algorithm}
\usepackage{algorithmic}
\usepackage{breqn}
\usepackage{amssymb}
\usepackage{pifont}
\newcommand{\cmark}{\ding{51}}
\newcommand{\xmark}{\ding{55}}
\floatstyle{plaintop}
\restylefloat{table}

\usepackage{fancyhdr, lipsum}

\makeatletter
\let\ps@plain\ps@fancy 
\makeatother

\begin{document}
%
\title{Truthful Market-based Trading of Cloud Resources with Reservation Price}


\author{\IEEEauthorblockN{Sergei Chichin, Quoc Bao Vo, Ryszard Kowalczyk}
\IEEEauthorblockA{Faculty of Information \& Communication Technologies\\
Swinburne University of Technology,\\ 
Melbourne, Australia\\
\{schichin, bvo, rkowalczyk\}@swin.edu.au}
}


%


\maketitle
\thispagestyle{fancy}

\begin{abstract}
With the rapidly growing demand for the cloud services, a need for efficient methods to trade computing resources increases. Commonly used fixed-price model is not always the best approach for trading cloud resources, because of its inflexible and static nature. Dynamic trading systems, which make use of market mechanisms, show promise for more efficient resource allocation and pricing in the cloud. However, most of the existing mechanisms ignore the seller's costs of providing the resources. In order to address it, we propose a single-sided market mechanism for trading virtual machine instances in the cloud, where the cloud provider can express the reservation prices for traded cloud services. We investigate the theoretical properties of the proposed mechanism and prove that it is truthful, i.e. the buyers do not have an incentive to lie about their true valuation of the resources. We perform extensive experiments in order to investigate the impact of the reserve price on the market outcome. Our experiments show that the proposed mechanism yields near-optimal allocations and has a low execution time.
\end{abstract}


%
\IEEEpeerreviewmaketitle

\vspace{-8px}

\section{Introduction}
\vspace{-3px}
The market of cloud services is rapidly growing. The number of enterprises and individuals, who outsource their workloads to the cloud providers, is increasing \cite{cloudgrowth}\cite{chichin2013}. Infrastructure as a Service (IaaS), which is one of the cloud provisioning models, allows an organization to outsource the computational resources needed to support "operation". Most commonly, IaaS is offered in the form of virtual machines (VM), which are characterized by a number of components, such as CPU, memory, storage, networks and other low level resources. The provisioned resource is typically charged on-demand or by subscription. Each traded VM type (i.e. instance type) has its own price, which is pre-defined by the cloud provider and does not change dynamically. Due to the inability of the price to adapt to the demand, the resource is often traded inefficiently \cite{bakos1998}\cite{cortese1998}. The market-based mechanisms for trading cloud resources start to be used by the cloud providers. An example is Amazon EC2 Spot Market \cite{amazonspot}, which enables an efficient provisioning of unutilized resource on EC2. While Amazon EC2 Spot Market conducts separate auctions for the VMs of different types (e.g. small, medium, large), we argue that allowing more complex user requests (than a single instance request) in a form of combinatorial bids is more efficient \cite{zaman2010} and more convenient for the users with complex requirements.

A market mechanism for trading cloud services must satisfy a number of requirements that are essential for the resource consumers and a cloud provider (we consider a problem with a single resource provider). The considered requirements and the motivation are provided below:

\vspace{-2px}
\begin{itemize}
\item 
\emph{Reserve price:} Cloud service provisioning incurs a certain cost. Selling goods at the price below that cost results in loss. This is the reason why the electronic markets consider the minimum seller price. For example, eBay has a reservation price \cite{eBayRP}. The cloud providers typically set minimum prices for their services. For example, a reservation price is one of the constraints in Amazon EC2 Spot Market\cite{amazonspot}. In our work we consider the reservation price for the traded resources when designing the market mechanism to enable the seller to gain a certain control over the established market prices and to avoid non-profitable trade.
\item 
\emph{Fast market clearance:} Cloud computing is a very dynamic environment with large amount of users. The market mechanism for cloud resources has to have a low computation complexity in order to be feasible in real cloud market.
\item 
\emph{Resource bundling:} Very often, when outsourcing IT infrastructure to the cloud, the buyers look for higher reliability, or wish to balance the traffic load. In such situation there is a need for a complex infrastructure, which is composed of a number of different VMs, that may be required, for example in different geographical regions or availability zones \cite{availzones}.
Resource bundling allows a more convenient way of accommodating such complex requests. First of all, it allows the cloud consumers to express their complex infrastructure requirements in a single request. Secondly, the budget constraint applies to a set of resources and doesn't have to be split across multiple requests. In this work we consider a combinatorial setting in order to address this issue. Combinatorial auctions, which involve trading of bundles are proven to be more efficient \cite{efficientcombauctions} and are successfully implemented for selling wireless spectrum or in transportation procurement \cite{combauctionapplication}\cite{combauctionapplication2}.
\end{itemize}

In addition we consider the following design desiderata:

\begin{itemize}
\item 
\emph{Allocative efficiency:} An allocative efficient mechanism is able to maximize the total utility (i.e. social welfare) of the market participants (buyers), given a particular input.
\item
\emph{Truthfulness:} 
A truthful mechanism ensures that the competition in the market is based on a single weakly dominant strategy, which is to reveal the true valuation for the requested resources. Truthful market mechanism ensures fairness because the buyers are not disadvantaged by telling the truth. Moreover, this market property significantly simplifies the strategy space of the buyers.
In this work we design a truthful market mechanism, which eliminates the buyer's strategic behavior, making truth telling a dominant buyer's strategy.
\end{itemize}

\emph{Our contribution:} We address the problem of VM provisioning and allocation in the cloud. We design a combinatorial market mechanism for a single resource provider, that allows the seller to express the minimum desirable price for traded resources (i.e. reserve price). We investigate the theoretical properties of the proposed market mechanism and prove that it is truthful, i.e. the buyer's dominant strategy is to reveal her true valuation for the resources. We evaluate the performance of the proposed mechanism in extensive simulation experiments. The experimental results show that the proposed allocation scheme achieves near optimal allocations and has a low execution time. We study the impact of various market conditions (e.g. resource over- or under-provisioning) on the market outcomes (e.g. resource utilization, generated revenue, etc.), and we define the regular rules for achieving various objectives, such as maximizing the expected revenue or resource utilization.

The rest of the paper is organized as follows. In Section \ref{literature}, we discuss the most relevant market-based approaches for provisioning IaaS. In Section \ref{problem}, we describe the problem of resource allocation in clouds. In Section \ref{mechanism}, we describe our mechanism and discuss its theoretical properties. Section \ref{experiments} is dedicated to the experimental evaluation of our proposed mechanism. In Section \ref{conclusion} we make several concluding remarks and discuss the future work.

\section{Related Work}
\label{literature}
Market-based approach in IaaS is attracting more attention in research community. We present the most closely related works in this domain.

Lehmann et al. \cite{lehmann2002} discuss a truthful market mechanism based on the greedy heuristics, which has become a fairly popular approximation allocation mechanism for combinatorial domains.
Zaman and Grosu \cite{zaman2010} and Nejad et al. \cite{lena2013} study combinatorial auction mechanisms for VM allocations in the cloud with the unique resource provider. They propose the greedy mechanisms, and a linear programming relaxation and randomized rounding mechanism. They prove that the greedy mechanisms are truthful and linear programming mechanism is truthful in expectation. Their work solves similar problem to ours, but they assume that there is no resource associated cost (and no resource reservation price), which is hardly realistic.
St{\"o}{\ss}er et al. \cite{stober2010} propose a knowledge-based continuous double auction mechanism to determine the prices of the future trades. They design an approximation mechanism based on greedy heuristic and analyze the strategic behavior in the market. Their proposed solution is not suitable for combinatorial requests and does not guarantee truthfulness.

Lampe et al. \cite{lampe2012} address a problem of concurrent pricing and distribution of VMs across physical machines with the objective of maximizing the seller's revenue. They propose an equilibrium price auction, where the demand is distributed among multiple types of resources, each with its own equilibrium price. Their approximation mechanism (with reservation price), while achieving near-optimal allocation, does not ensure truthfulness, which is a market property that we maintain in our proposed mechanism.

Fujiwara \cite{fujiwara} addresses a combinatorial problem of trading cloud services in different timeslots. He designs an optimal combinatorial mechanism using linear mixed integer program for resource allocation and uses k-pricing scheme to determine the prices. The developed mechanism allows untruthful bidding and tends to be computationally hard. Our proposed approximation solution solves these two issues.

Xingwei et al. \cite{xing-wei2012} consider a periodical combinatorial auction with a single seller. They propose to determine the resource prices based on limited English combinatorial auction model, and to optimally allocate resources in different timeframes based on genetic algorithm. Their proposed mechanism does not guarantee truthfulness.

Zhao et al. \cite{CloudBay} describe a marketplace for trading cloud resources based on e-Bay-like transaction model that supports different services with different level of job priorities. Their marketplace contains an auction system that encourages truthful bidding. Unlike their work, we consider combinatorial market setting with a unique seller without the requirement to prioritize the requests.

Roovers \cite{roovers} investigates a problem of double-sided allocation in clouds. He shows that continuous double auction is not feasible in the considered market setting and propose continuous reverse auction to address the problem. Unlike that work, we consider combinatorial setting and design a truthful periodic market mechanism, because periodic mechanisms are proved to generate more efficient allocations compared to continuous market mechanisms.

All of the studies listed above propose the market mechanisms for trading cloud resources, but none of them design a truthful combinatorial market mechanism with the reservation price. As was shown before, truthfulness, reservation price and bundling are very important practical requirements for the cloud market. One of the key differences of our work is that we design a truthful combinatorial auction with reservation price and investigate its performance, including the impact of reservation price on the resource utilization, generated revenue, offered discount, etc. To the best of our knowledge there is no work presented in literature that investigates mechanism's truthfulness with reservation price.

\section{Problem Formulation}
\label{problem}

We formulate a problem of cloud resource allocation. All the notations are summarized in Table \ref{Notation}.

\emph{Seller:} A cloud provider (i.e. seller) has a set of physical resources (e.g. CPU, RAM, storage) that she wishes to sell to the buyers (i.e. users, bidders) in the form of Virtual Machines (VMs).

We assume that the cloud provider offers $k$ different instance types and each contains different amount of physical components. For example, Amazon EC2 offers small, medium, and large instances with their own pre-defined characteristics in terms of CPU, RAM, and storage \cite{amazoninst}.
A seller declares her reserve prices for the VMs, (i.e. offer prices), which are characterized by vector: $\langle o_1,...,o_k\rangle $. These are the minimum prices that the seller wants for her resources and she would not sell for the price below. The seller specifies the amounts of supplied VMs of each type, which are defined as $\langle s_1,..., s_k \rangle$.  
The seller's ask is expressed as follows: $a=(\langle s_1,..., s_k \rangle, \langle o_1,..., o_k\rangle)$.
For example, $a = (\langle90, 50, 20 \rangle, \langle \$0.06, \$0.12, \$0.24 \rangle)$ supplies 90, 50 and 20 VMs of three different types (e.g. small, medium and large) and offers them at least for the specified reserve prices: \$0.06, \$0.12 and \$0.24 per item, respectively.

\emph{Buyers:} 
We assume that $n$ potential buyers (i.e. bidders) participate in the auction $U=\{ u_1,...,u_n \}$. All of the bidders have various cloud service requirements and the budget constraints. Thus, the bid is combinatorial and includes the information about the set of required resources (bundle) $d_j=\langle r_{1j},...,r_{kj}\rangle$, and the willingness to pay for the bundle (valuation) $v_j\ge0$. We assume that each buyer submits a single bid.
We consider a problem, where all bidders in the auction are single-minded. A single-minded bidder has only two possible states:  either she obtains the entire requested bundle or any superset of it (for the price within her valuation) or she values the resource at 0 (her utility for partial request fulfillment is 0). 
A buyer's bid is expressed as: $b_j=(\langle r_{1j},...,r_{kj}\rangle ,v_j )$.
For example, $b_x = (\langle2, 0, 1 \rangle, \$5)$ request two VMs of type 1 and one VM of type 3 and the declared value for the bundle is \$5. 

\emph{Virtual Resource Allocation Problem:}
The goal of the virtual resource allocation problem is, given a set of buyers $U$ and their bids $b_j$, to determine the set of winners $W\subseteq U$ and the prices, $p_j$, that they pay to the seller, such that:
\begin{equation}
	\sum_{j:u_j\in W} r_{ij} \le s_i, i=1,\ldots,k
\end{equation}
\begin{equation}
	\textit{\^{o}}(d_j) \le p_j \le v_j \mbox{, if } u_j \in W 
\end{equation}
\begin{equation}
	p_j=0 \mbox{, if } u_j \notin W
\end{equation}

In the constraint (1) we ensure that the traded resource capacity is not exceeded; referred to as Available Resource Constraint (ARC). The constraint (2) confirms that the winning user pays at most her declared valuation, and at least the price that the seller wants for the requested bundle of resource ($d_j$). We define the bundle-specific reserve price as $\textit{\^{o}}(d_j) = \sum_{i=1}^{k} r_{ij} o_i$, which is the weighted sum of reserve prices for all the requested resources in the bundle. We refer to $v_j \ge \textit{\^{o}}(d_j)$ as Reserve Price Constraint (RPC). The constraint (3) implies that the losing bidders does not pay anything.
The standard objective in combinatorial auction design is to maximize the sum of the winning bidder's valuations \cite{nisan2007}. The objective in our problem is to maximize $V=\sum_{j:u_j\in W} v_j$, subject to the constraints (1)-(3).

\section{Market Mechanism}
\label{mechanism}

\subsection{Mechanism Design Framework}

The described allocation problem can be considered as a multi-dimensional knapsack problem (MKP) with an additional constraints (2)-(3). However as MKP is NP-hard \cite{martello1990}
we propose a mechanism with approximate solution.

\begin{table}[t]
\caption{Notation}
\centering
\begin{tabular}{c|l}
\hline
$a$				& The seller's \underline{ask}							\\
$k$				& Number of VM types proposed by the seller					\\
$s_i$				& \underline{Supplied} amount of VMs of type $i \le k$			\\
$o_i$				& Reserve (\underline{offer}) price for the VM of type $i \le k$		\\
$U$ 				& Set of buyers (\underline{users}) \{$u_{1}, ..., u_{n}$\}			\\
$b_j$				& \underline{Bid} of the buyer $u_j$						\\
$r_{ij}$			& The amount of VMs (\underline{resource}) of type $i \le k$ 	 	\\
				& requested by the user $u_j$							\\
$d_j$				& The bundle of VMs requested (\underline{demanded}) by $u_j$ 		\\
$v_j$				& Declared \underline{valuation} of the user $u_j$ \\
$p_j$				& \underline{Price} that the user $u_j$ pays					\\
$W$				& Set of \underline{winning} buyers $W \subseteq U$			\\
\hline
\end{tabular}
\label{Notation}
\vspace{-15px}
\end{table}

The goal is to design a truthful mechanism with the reserve price constraint, that maximizes the social welfare $V$. Each bidder in the market has her own true type, which is private information. By type we denote a function, which determines the bidder's true valuation for the bundle of resources $\textit{\^{v}}_j(d_j)$. Note that the true valuation for the bundle, determined by function $\textit{\^{v}}_j(d_j)$, may be different from the declared value in the bid $v_j$.

\emph{Definition 1 (Truthfulness):} A truthful mechanism makes reporting a true user type a dominant strategy, i.e. the bidders maximize their utility by truthful reporting regardless of the other users' reported bids. For a market mechanism to be truthful, its allocation scheme must be monotone and the pricing must be based on the critical value.

\emph{Definition 2 (Monotonicity):} Monotonicity is a property, which implies that the winning bid cannot lose by offering more money for fewer goods, and the losing bid cannot win by offering less money for more goods. Thus, if a bid $b_1$ won, changing the bundle $d_1 \rightarrow d_{2}\mbox{, so that } d_{2} \subseteq d_1$\footnote {Given two bundles $d_1=\langle r_{11},...,r_{k1}\rangle$ and $d_2=\langle r_{12},...,r_{k2}\rangle$, we write $d_1 \subseteq d_2$ if $r_{11}\le r_{12} \&\ldots\&r_{k1}\le r_{k2}.$} or increasing the reported valuation $v_1 \rightarrow v_{2}\mbox{, so that } v_{2} \ge v_1$ would result in the new bid being also allocated. The symmetrical situation with the losing bid must also be true.

\emph{Definition 3 (Critical value pricing):} When a mechanism has a monotone allocation scheme, each bid would have a unique value $v_j^{cr}$, called critical value, such that the reported valuation above this value would result in the bid being granted the requested resource (if $v_j \ge v_{j}^{cr} \Rightarrow u_j \in W$), but any valuation below it would make the bidder lose (if $v_j < v_{j}^{cr} \Rightarrow u_j \notin W$). A losing bid must pay nothing. The critical-value pricing is determined as follows:
\begin{equation}
	p_j=
	\begin{cases} 
		v_j^{cr}, & \mbox{if } u_j \in W \\
		0, & \mbox{otherwise}
	\end{cases}
\end{equation}

\subsection{Truthful Greedy-RP Mechanism}

We consider the Greedy heuristic \cite{lehmann2002} as the basis for our mechanism which is called Greedy-RP.
A greedy mechanism consists of two steps: allocation and pricing. During the allocation phase, the mechanism decides who are going to be the winners and receive the resources. The pricing phase determines the prices that the bidders have to pay.

\emph{Allocation Scheme: }
Greedy-RP allocation scheme is given in Algorithm \ref{allocation}. The algorithm receives the seller's offer, which is a vector of resource capacities and reserve prices, and the vector of bids. The output of the allocation scheme is a set of determined winning users.
 
The idea of greedy allocation is to rank the bids based on simple value (i.e. bid density). Generally, when allocating homogeneous goods this value is a price per item. As we consider combinatorial bids with the VMs of different types, we need to know the relative relation (i.e. relativity) between these resources in order to determine the bid density. The relative relation  is defined by vector $\langle f_1,\dots,f_k\rangle$, where $f_i>0$ and $f_1\le f_2\le\dots\le f_k$. As the reserve price is often a reflection of resource cost, it could be a good measure for the resource relative relation, because it takes into account all the VM components (VM cost is a sum of individual VM components costs). When we know the relativity between the resources of different types, we can calculate the total weight of the requested bundle as a weighted sum of the resources in the bid $\textit{\^{d}}(d_j)$ (line 4) and re-order the users in non-increasing order of their bid densities (line 5), which are defined as follows:
\begin{equation}
	\textit{\^{d}}(d_j) = \sum_{i=1}^k r_{ij} f_i
\end{equation}
\begin{equation}
	e(b_j)=\frac{v_j}{|\textit{\^{d}}(d_j)|^q}, \mbox{ where } q>0
\end{equation}

\vspace{2px}
Please, note that the bid density form with the parameter $q$ was proposed by Lehmann \cite{lehmann2002}, and changing the parameter has an impact on the efficiency of resource allocation. We design a market mechanism with the bid density that considers $q$ because changing this parameter according to the specific market conditions can allow to flexibly improve the efficiency of the allocation, as was investigated by Lehmann in a single-item market mechanism setting. The experimental investigation of this impact in a multiple item auction is out of scope of this work and will be investigated in future.

After sorting the users, 
the mechanism greedily allocates the resources to them in the sorted order, subject to the ARC (1) and RPC (2) (lines 6-10). If both constraints are satisfied, the user becomes the winner (line 8) and she will receive exactly the requested bundle (line 7); otherwise the bid is denied.

\setlength{\textfloatsep}{5pt}
\begin{algorithm}[t!]
\caption{Greedy-RP Allocation Scheme (GRP-A)}
\begin{algorithmic}[1]
\STATE {\bf Input:} $a=(\langle s_1,..., s_k \rangle, \langle o_1,..., o_k\rangle)$; ask, which contains a vector of resource capacities and reserve prices
\STATE {\bf Input:} $B=(b_1, \dots, b_n)$; vector of bids (bundle, value)
\STATE $W \leftarrow \emptyset$
\STATE Compute $\textit{\^{d}}(d_j)$ according to (5)
\STATE Re-order users such that: $e(b_1) \ge e(b_2) \ge \dots \ge e(b_n)$ (6)
\FOR {$j=1,\ldots,n$}
	\IF{$\mbox{for all } i=1,\ldots,k, r_{ij}+\sum_{j': u_{j'} \in W} r_{ij'} \le s_i$ \\
		$\mbox{    }\mbox{    }\mbox{    }\mbox{    }\mbox{    }\mbox{    }$     \AND $\textit{\^{o}}(d_j) \le v_j$}
		\STATE $W \leftarrow W \cup u_j$
	\ENDIF
\ENDFOR
\STATE {\bf Output:} W
\end{algorithmic}
\label{allocation}
\end{algorithm}

\newtheorem{lma}{Lemma}
\begin{lma}
The Greedy-RP Allocation Scheme is monotone.
\end{lma}
\noindent
\begin{proof}
To prove bid monotonicity we consider two bids: $b_1=(d_1,v_1)$ and $b_2=(d_2,v_2)$, where $v_1 \ge v_2$ and $d_1 \subseteq d_2$. We will show that the winning bid cannot lose by offering more money for fewer goods and the losing bid cannot win by paying less for more goods.

If we calculate the bid densities based on presented form $e(b)= \frac{v}{|s|^q}$, where $q>0$. Since $d_1 \subseteq d_2$, then $0 \le |s_1| \le |s_2|$. As $q>0$, then $|s_1|^q \le |s_2|^q$. Since $v_1 \ge v_2$, we will get: $e(b_1) \ge e(b_2)$. Thus, by changing $(d_1,v_1) \rightarrow (d_2,v_2)$, the sorted list of bids may differ only in the bid $b_2$ being moved down in the sorted list.
We call $L_1$ the sorted list with the bid $b_1$, and $L_2$ the list with the bid $b_2$. If $b_1$ was denied in $L_1$ there are two possible reasons for this:
\begin{itemize}
\item	\emph{ARC:} there is another bid $b_c$ which is ordered before the bid $b_1$ and there was not enough resource for $b_1$. Then, $b_c$ would also be ordered before bid $b_2$ in the list $L_2$ (because $e(b_1) \ge e(b_2)$) and, as $d_1 \subseteq d_2$, the bid $b_2$ would be denied as well.
\item \emph{RPC:} if the bid $b_1$ was denied because of violating the RPC, it means that $v_1 < \textit{\^{o}}(d_j)$, where $\textit{\^{o}}(d_j) = \sum_{i=1}^{k} r_{ij} o_i$.
Since $d_1 \subseteq d_2$, we get $r_{11} \le r_{12} \& \ldots \& r_{k1} \le r_{k2}$. Thus $\sum_{i=1}^k r_{i1} o_i \le \sum_{i=1}^k r_{i2} o_i$, which is $\textit{\^{o}}(d_1) \le \textit{\^{o}}(d_2)$. Since $v_2 \le v_1 < \textit{\^{o}}(d_1) \le \textit{\^{o}}(d_2)$, we get $v_2<\textit{\^{o}}(d_2)$. It means that the bid $b_2$ would not satisfy the RPC neither.
\end{itemize}
Similarly, the opposite is true: if the bid $b_2$ was granted in the list $L_2$, the bid $b_1$ will be granted in the list $L_1$. Hence, the proposed allocation scheme is monotone.
\end{proof}

\emph{Pricing Scheme: } The pricing scheme receives the seller's offer, vector of bids and the set of winning bidders as its input. It aims to determine the prices that the market participants have to pay. The payment vector $P$ is the outcome of the pricing scheme. 
We aim to define a general pricing mechanism for any bid density of the specified form, which would determine the critical-value payments for the winning bids in $W$. Critical-value price is the minimum possible price that the winner could have proposed for the bundle in order to win the auction. 

The payment scheme determines the payments for the winning bids in order (line 4). In greedy market mechanisms the bid density is used to determine the final prices that the bidders pay. In our mechanism the final determined price that the winner pays is either based on the density of the first losing competitor bid $e_j^{comp}$ (also referred to as competitor bid density), or on the bid-specific reserve density $e_{res}(b_j)$. First of all, the mechanism finds the first losing competitor bid. The first losing competitor bid for the bid $b_j$ is the bid that lost, but would have been granted if $b_j$ did not participate in the market. In order to find it, the mechanism constructs a new market setting, which excludes the current bid and runs the allocation scheme with the new setting (lines 5-6). After that, the mechanism selects only the new winners compared to the initial allocation (line 7) and determines the one with the highest bid density, which would be the first losing bid (lines 8-12). Such a bid may not exist and even if it exists, its density may be lower than the required density needed for the calculated price to satisfy the reserve price constraint (Const. 2). It happens because of the non-linear nature of the bid density function, when the parameter $q\neq 1$. In order to address that, we introduce the bid-specific reserve density:
\vspace{5px}
$$
e_{res}(b_j) = \frac{\textit{\^{o}}(d_j)}{|\textit{\^{d}}(d_j)|^q}
$$

It is the minimum possible density, which when applied to the bundle $d_j$ would generate the price that would satisfy the constraint RPC.

We determine the critical density $e_j^{cr}$ of the bid $b_j$ by selecting the highest density value between the first losing competitor bid and the reserve bid density (line13). 
We determine the final price of the bid $b_j$ based on the critical bid density $e_j^{cr}$, if the user was the winner (line 14), and the final price is set to 0 for the losing users (line 16-18).

\setlength{\textfloatsep}{10pt}
\begin{algorithm}[t!]
\caption{Greedy-RP Pricing Scheme (GRP-P)}
\begin{algorithmic}[1]
\STATE {\bf Input:} $a=(\langle s_1,..., s_k \rangle, \langle o_1,..., o_k\rangle)$; ask, which contains a vector of resource capacities and reserve prices
\STATE {\bf Input:} $B=(b_1, \dots, b_n)$; vector of bids (bundle, value)
\STATE {\bf Input:} $W$; winning users
\FORALL {$u_j \in W$}
	\STATE $B' \leftarrow B \setminus \{b_j\}$
	\STATE $W' \leftarrow \mbox{ GRP-A(a, B')}$
	\STATE $W'_j \leftarrow W' \setminus W$
	\IF {$W'_j \neq \emptyset$}
		\STATE $e_j^{comp} \leftarrow \max_{b_h : u_h \in W'_j}(e(b_h))$
	\ELSE
		\STATE $e_j^{comp} \leftarrow 0$
	\ENDIF
	\STATE $e_j^{cr} \leftarrow \max(e_j^{comp}, e_{res}(b_j))$
	\STATE $p_j \leftarrow (e_j^{cr})|\textit{\^{d}}(d_j)|^q$
\ENDFOR
\FORALL {$u_j \notin W$}
	\STATE $p_j \leftarrow \emptyset$
\ENDFOR
\STATE {\bf Output:} $P=(p_1,\dots,p_n)$; the payment vector
\end{algorithmic}
\label{pricing}
\end{algorithm}

\begin{lma}
The Greedy-RP Pricing determines critical-value payment.
\end{lma}
\noindent
\begin{proof}
Critical-value price is the lowest possible price that a winning bidder could have proposed for the requested bundle in order to still be the winner.
In our mechanism the price $p_j$ a winning bidder $u_j$ has to pay is based on the critical density, which may either be the first losing bid density $e_j^{comp}$ or 
the bid-specific reserve density $e_{res}(b_j)$. Thus, there are two possible cases:
\begin{itemize}
\item	\emph{$p_j$ is based on $e_j^{comp}$: } This density is the lowest possible density for the bid to be competitive enough to win the auction, if it satisfies the constraint RPC. Any bid with a density less than $e_j^{comp}$ would result in another competitor bid being granted instead. Hence the price $p_j$ generated based on $e_j^{comp}$ is the critical-value.
\item \emph{$p_j$ is based on $e_{res}(b_j)$:} When there are no losing bids to a bid $b_j$ or when the bid density of the highest losing bid does not meet the reserve density $e_j^{comp}<e_{res}(b_j)$, the reserve bid density is used to calculate the price $p_j$. Any bid with a density less than $e_{res}(b_j)$ would result in the bid being refused by the constraint RPC. Thus, this density generates the critical-value price.
\end{itemize}
We can see that our proposed pricing scheme determines a critical value price for the winners and does not charge the lost bidders (lines 16-18), which corresponds to the critical value pricing definition (Eq. 4).
\end{proof}

\newtheorem{thm}{Theorem}
\begin{thm}
The Greedy-RP Mechanism is truthful for single-minded bidders.
\end{thm}
\noindent
\begin{proof}
We prove that the proposed mechanism contains four properties introduced by Lehmann et al. \cite{lehmann2002} that are shown to be sufficient for the mechanism to be truthful.

\emph{Exactness:} 
Given an order of the bidders according to their (non-increasing) bid densities, our allocation scheme satisfies exactness because the bidders 
either receive the exact requested bundle or don't receive anything.

\emph{Participation:} This property requires that a losing bidder pays nothing. In our mechanism, a bid $b_j$ is either granted or denied, and denied bidder pays nothing.

\emph{Monotonicity and Critical-value:} See Lemma 1 and 2.


As our proposed mechanism maintains all four properties, it is proved to be truthful for single-minded bidders.
\end{proof}

\emph{Computational complexity: } The allocation scheme in the proposed mechanism runs in polynomial time in the size of its input. The major computation is in order to sort the list of users. The sorting phase runs in $O(|N|\log|N|)$ time. The constraints verification and the actual resource allocation computes in $O(N)$. The pricing scheme runs as maximum N times and invokes winner determination each time, thus its computation complexity is $O(|N^2|\log|N|)$. 

\emph{Budget-balance: } In our proposed mechanism the payments made by the buyers are the payments to the seller, which makes our mechanism weakly budget balanced.

\emph{Individual Rationality: } The constraints (2) and (3), considered in our proposed mechanism, ensure individual rationality.

\subsection{Mechanism example}
In order to illustrate the way the mechanism works, we consider a simple example. We have five bids participating in the auction as depicted in the Table \ref{GRPExample}, and the Ask $a=(\langle 4, 4 \rangle , \langle \$8.0, \$16.0 \rangle)$. We assume that the relative relationship between the resources is based on the ratio between resource reserve prices. As $VM_2$ is twice more expensive, we assume that its relativity is $\langle 1, 2\rangle$. In our example, we consider two different bid densities, when $q=1$ - linear density, and when $q=\frac{1}{2}$ - square root density.

\begin{table}[t!]
\caption{Greedy-RP Example: Bids}
\centering
\begin{tabular}{c|| c c| c|| c| c c| c}
\hline\hline
$b_j$ & \multicolumn{2}{|c|}{$d_j$} & $v_j$ & $\textit{\^{d}}(d_j)$ & \multicolumn{2}{|c|}{Density} & Res. \\
 & $VM_1$ & $VM_2$ &  & & $q$=1 & $q$=$\frac{1}{2}$ & Price \\ [0.5ex]
\hline
$b_1$		& 1 		& 	0	& 	\$10 	& 1		&	10.0		&	10.0	&	\$8 \cmark		\\
$b_2$		& 0 		& 	1	& 	\$19 	& 2		&	9.5		&	13.4	&	\$16 \cmark	 	\\
$b_3$		& 2 		& 	2	& 	\$59 	& 6		&	9.8		&	24.1	&	\$48 \cmark		\\
$b_4$		& 3 		& 	1	& 	\$51 	& 5		&	10.2		&	22.8	&	\$40 \cmark		\\
$b_5$		& 1 		& 	1	& 	\$23 	& 3		&	7.7		&	13.3	&	\$24 \xmark		\\ [1ex]
\hline
\end{tabular}
\label{GRPExample}
\end{table}

\begin{table}[t!]
\caption{Greedy-RP Example:Allocation Phase}
\centering
\begin{tabular}{c c c c}
\hline\hline
Step & & \multicolumn{2}{c}{Allocation Results}\\
& & $q=1$ & $q=\frac{1}{2}$ \\ [0.5ex]
\hline
1&Sorted order		& $b_4, b_1, b_3, b_2, b_5$	&	$b_3, b_4, b_2, b_5, b_1$	\\
2&RPC			& $b_4, b_1, b_3, b_2$		&	$b_3, b_4, b_2, b_1$		\\
3&ARC			& $b_4, b_1, b_2$			&	$b_3, b_2$				\\ [1ex]
\hline
\end{tabular}
\label{AllocationExample}
\end{table}

\begin{table}[t!]
\caption{Greedy-RP Example: Pricing Phase}
\centering
\begin{tabular}{c | c c | c | c}
\hline\hline
 $b_j$& $e_{j}^{comp}$ & $e_{res}(b_j)$ & $e_{j}^{cr}$ & $p_j$ \\
 \multicolumn{5}{c}{$q=1$} \\ [0.5ex]
\hline
$b_4$	& $e(b_3)=9.8$	&	$8.0$	&	$9.8$		&	$9.8*5=\$49$	\\
$b_1$	& no Bid 		&	$8.0$	&	$8.0$		&	$8.0*1=\$8$	\\
$b_2$	& no Bid 		&	$8.0$	&	$8.0$		&	$8.0*2=\$16$	\\ [1ex]
\hline
 \multicolumn{5}{c}{$q=\frac{1}{2}$} \\ [0.5ex]
\hline
$b_3$	& $e(b_1)=10.0$ 	&	$19.6$&	$19.6$	&	$19.6*6^{\frac{1}{2}}=\$48$	\\
$b_2$	& $e(b_1)=10.0$ 	&	$11.3$&	$11.3$	&	$11.3*2^{\frac{1}{2}}=\$16$	\\ [1ex]
\hline
\end{tabular}
\label{PricingExample}
\vspace{-5px}
\end{table}

We can see that, based on the selected bid density type, the mechanism favors different types of bids (Table \ref{AllocationExample}): we obtain different bid orders in two considered cases. In general, when $q$ is low, the bid that requests more resource is favored, which may have a positive impact on the utilization. The sorted bids order is not critical when verifying the RPC. All the bids with the valuation below the required minimum are denied. The ARC considers the bids in the sorted order, and the outcomes may be different with different bid density forms. As a result of allocation scheme, the winners are determined - 3 winners when $q=1: (b_4, b_1, b_2)$, and 2 winners when $q=\frac{1}{2}: (b_3, b_2)$.

During the pricing phase (Table \ref{PricingExample}), the bid-specific reserve density serves as a verifier for the reserve price constraint. When the bid density form is linear and there is no first losing bid (competitor bid), the bid-specific reserve density sets the final price for the bid to the value of reserve price for the bundle. When the bid density form is not linear ($0<q<1$), the price generated by the density of the first losing bid is often below the reserve price for the bundle. The bid-specific reserve density ensures that the RPC is not violated. For example, if the density of the competitor bid $b_1$ was used to determine the final price for the bid $b_3$ ($q=\frac{1}{2}$), it would be $10*6^{1/2}=\$24.5$, which is almost half of the initial reserve price of $\$48$.

\section{Experiments}
\label{experiments}

In this section, we investigate experimentally the proposed mechanism's behavior with various market settings, involving resource under-provisioning (UP) when supply $<$ demand, resource over-provisioning (OP) when supply $>$ demand, and resource exact-provisioning (EP) when supply $=$ demand. 

First, we demonstrate truthfulness of the proposed mechanism. After that, we conduct two experiments. The aim of the first experiment is to investigate the proposed mechanism's behavior. 
In the second experiment, we examine the allocation performance of the proposed mechanism in comparison with other allocation schemes for MKP.

\begin{table}[t]
\caption{True valuation bids solved}
\centering
\begin{tabular}{c | c | c | c | c | c | c | c}
\hline \hline
Bid 		& $d_j$ 			& $v_j$	& $e(b_j)$ 	& W 		& $e_{j}^{comp}$ 	& $\textit{\^{o}}(d_j)$	& $p_j$ 	\\
\hline
$b_1$		& $\langle1,2,1\rangle$	& $\$7.2$	& 2		& {\bf Yes}	& 5.4			& 3.6		& {\bf 5.4}	\\
$b_2$		& $\langle0,1,3\rangle$	& $\$14$	& 2.5		& {\bf Yes}	& 8.4			& 5.6		& {\bf 8.4}	\\
$b_3$		& $\langle1,0,1\rangle$	& $\$3$	& 1.5		& {\bf No}	& -			& 2.0		& {\bf -}	\\
 [1ex]
\hline
\end{tabular}
\label{truthexp}
\end{table}

\begin{table}[t]
\caption{Untruthful bidding scenarios (for bid $b_2$)}
\centering
\begin{tabular}{c | c | c | c | c | c | c}
\hline \hline
N 		& $d_{2'}$ 			& $v_{2'}$	 	& Scenario			& W 	& $p_{2'}$	& Util. 	\\
\hline
1		& $\langle$0,1,3$\rangle$	& 14	& $v_{2'}$=$v_2$, $d_{2'}$=$d_2$		& {\bf Yes}	& 8.4		& {\bf 5.6}	\\
2		& $\langle$0,1,3$\rangle$	& 18	& $v_{2'}>v_2$, $d_{2'}$=$d_2$		& {\bf Yes}	& 8.4		& {\bf 5.6}	\\
3		& $\langle$0,1,3$\rangle$	& 10	& $v_{2'}<v_2$, $d_{2'}$=$d_2$		& {\bf Yes}	& 8.4		& {\bf 5.6}	\\
4		& $\langle$0,1,3$\rangle$	& 6 	& $v_{2'}\ll v_2$, $d_{2'}$=$d_2$		& {\bf No}	& 0.0		& {\bf 0.0}	\\
5		& $\langle$1,1,3$\rangle$	& 14	& $v_{2'}$=$v_2$, $d_{2'}\supset d_2$	& {\bf Yes}	& 9.0		& {\bf 5.0}	\\
6		& $\langle$0,1,6$\rangle$	& 14	& $v_{2'}$=$v_2$, $d_{2'}\supset d_2$	& {\bf No}	& 0.0		& {\bf 0.0}	\\ 
[1ex]
\hline
\end{tabular}
\label{truthlie}
\vspace{-5px}
\end{table}

\subsection{Greedy-RP Truthfulness example:} We experimentally investigate truthfulness of our proposed Greedy-RP mechanism. In order to analyze the impact of untruthful bidding and to show that our mechanism is robust against the strategic manipulation by the buyers, we consider the example of three truly declared bids: $b_1=(\langle1,2,1\rangle, \$7.2)$,  $b_2=(\langle0,1,3\rangle, \$14)$, $b_3=(\langle1,0,1\rangle, \$3)$. The seller's ask supplies four resource units of each type with the reserve prices as follows: $a=(\langle4,4,4\rangle,\langle\$0.4, \$0.8, \$1.6\rangle)$. We assume that the relative relation between the instance types is based on the reserve prices. We also select linear bid density form ($q=1$) for simplicity.
The Greedy-RP determines the solution as depicted in the Table \ref{truthexp}. The bids $b_1$ and $b_2$ are allocated the resource and the determined prices are \$5.4 and \$8.4 respectively.

We suppose that the user with the bid $b_2$ lies and declares different untruthful types (lies about her valuation and bundle). The market outcome depends on the declared values of $d_{2'}$ and $v_{2'}$. We consider six different scenarios, which are depicted in the Table \ref{truthlie}. We determine the user's utility as the difference between the true type valuation (\$14) and the final determined price ($p_j$). In case 1, the buyer states her true bundle and valuation. In case 2, the buyer increases the valuation of the bid, but the determined price and the utility remains the same as in the true declaration because the price is a critical-value. In case 3, the valuation of the bid $b_2$ is reduced, but it is still above the critical value price ($\$10.0>\$8.4$), which allows the bid to win. The buyer's utility is the same as when the true type was specified because the determined price is also the same. In case 4, the declared valuation is below the critical value price ($\$6.0<\$8.4$), which is the reason for the bid to be denied and utility drop to 0. In case 5, the bid requests bigger bundle than in the true declaration. The bid is allocated, however the critical-value price increases, which has a negative impact on the buyer's utility. In case 6, the requested bundle is too large for the bid to be allocated, which results in 0 utility.
We can see that in all the cases, the user cannot increase her utility by untruthful bidding, which confirms mechanism's truthfulness.

\subsection{Experiment 1 (Greedy-RP Analysis):}  In this experiment, we examine our proposed mechanism (Greedy-RP). The aim of this experiment is to investigate how various market inputs influence the cloud resource utilization, social welfare (i.e. total offered discount), generated revenue and the total resource cost.

\emph{Experimental Setup:} Our experimental setup is depicted in the Table \ref{ExpSetup2}. In this work we consider linear bid density with $q=1$. The experimental investigation of various bid densities and its impact on the resource allocation is out of scope of this work and will be investigated in the future.

We assume that there are 50 buyers participating in the market, which is not large enough to result in extremely long computation time and is not too small to affect our experimental precision. The number of bidders doesn't have an impact on the market outcome in our experiments because we experiment with different supply values. Inspired by a real-world example, where the cloud providers estimate their possible gain or loss from resource over-provisioning or under-provisioning, we first establish the demand and generate different levels of supply (as a percentage of demand) in order to examine its impact.

As there is no publicly available information about the cloud buyers' requests, we achieve the variability in the market input by randomly generating the requested bundles of resources and valuations based on normal distribution functions (DF). We construct the distribution functions in the way that 99.73\% of the randomly generated values lie in the specific range. We regenerate all the values that are out of the required range (0.27\%). When generating the resource bundle, we apply the same DF for each resource type (RT), which are distributed over the range [0, 5]. The valuation of the bid is generated as a random value ($\textit{\={v}}_j$) for the smallest unit of resource ($o_1$) and is scaled according to the weighted amount of resource in the bundle ($\textit{\^{d}}(d_j)$).

We experiment with three different numbers of offered resource types ($k$). In order to define the relative relation between the VMs of different types we consider the prices of small, medium and large instances from Amazon EC2 \cite{amazoninst}, which scale based on such ratio 1:2:4 (rounded values).
We construct five provisioning levels based on the demand for each resource type ($5^k$ cases) in order to investigate the impact of different supply-demand conditions on the market outcome. We vary the reservation prices ($o_i$) within the valuation's range $\textit{\={v}}_j \in [0,1]$ (scaled according to $f_i$) in order to examine the impact of varying reserve price on the market outcome. It should also be noted that when $RP=0.0$, our mechanism performs as CA-Greedy\cite{zaman2010}, which allows us to compare those two different mechanisms in different settings.

Overall, we construct 1550 different market settings. We repeat each setting 1000 times and calculate the average values for the resource utilisation, generated social welfare (i.e. total buyers' utility), revenue, amount of sold and not sold resource. 

\begin{figure*}[h!]
\setlength{\belowcaptionskip}{-15pt}
\begin{center}
	\subfloat[Avg. Res. Utilization: RP=0.1] {
		\includegraphics[width=0.24\textwidth]{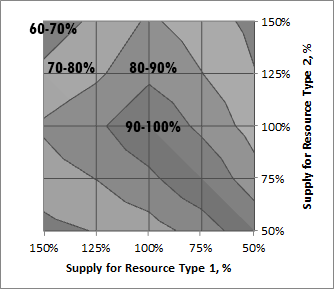} \label{u1}
	}
	\subfloat[Avg. Res. Utilization: RP=0.2] {
		\includegraphics[width=0.24\textwidth]{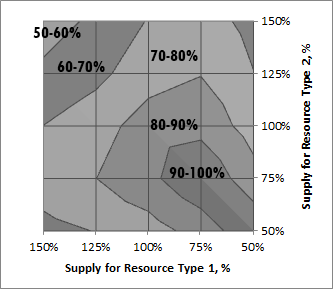} \label{u2}
	}
	\subfloat[Avg. Res. Utilization: RP=0.3] {
		\includegraphics[width=0.24\textwidth]{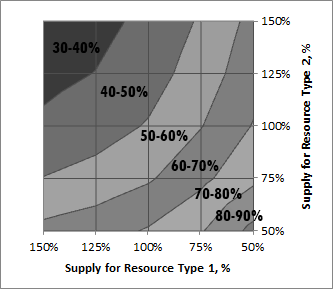} \label{u3}
	}
	\subfloat[Avg. Res. Utilization: RP=0.4] {
		\includegraphics[width=0.24\textwidth]{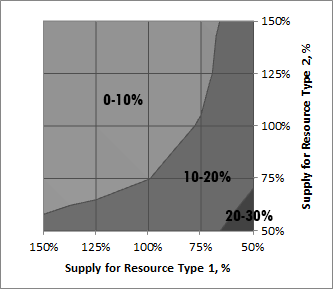} \label{u4}
	}

	\subfloat[Revenue: High UP (Supply from 50\%)] {
		\includegraphics[width=0.32\textwidth]{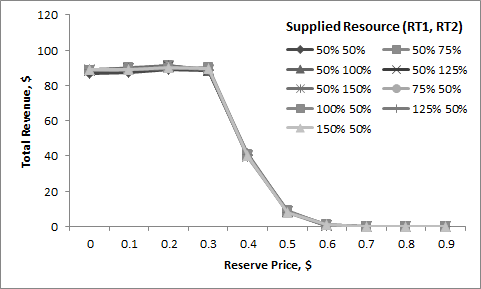} \label{rhUP}
	}
	\subfloat[Revenue: Low UP (Supply from 75\%)] {
		\includegraphics[width=0.32\textwidth]{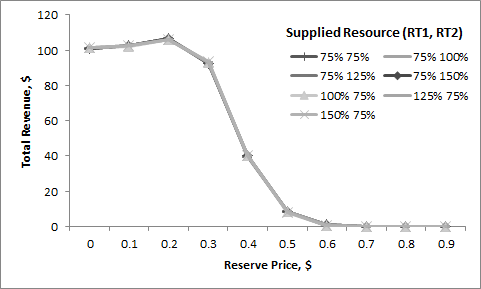} \label{rlUP}
	}
	\subfloat[Revenue: OP (Supply $>=100\%$)] {
		\includegraphics[width=0.32\textwidth]{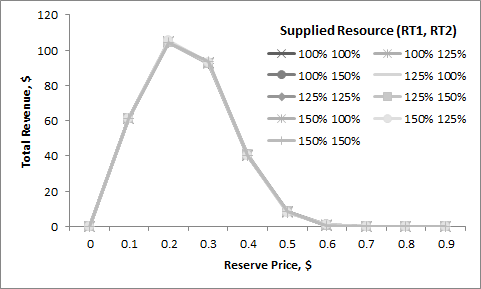} \label{rOP}
	}

	\subfloat[Buyer's Utility (Social Welfare): High UP (Supply from 50\%)] {
		\includegraphics[width=0.32\textwidth]{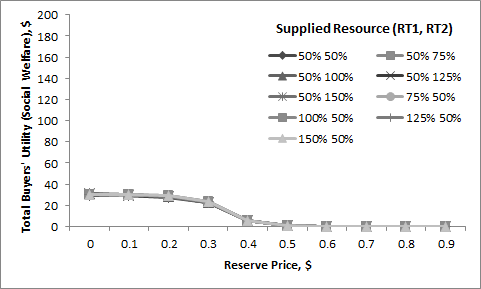} \label{buhUP}
	}
	\subfloat[Buyer's Utility (Social Welfare): Low UP (Supply from 75\%)] {
		\includegraphics[width=0.32\textwidth]{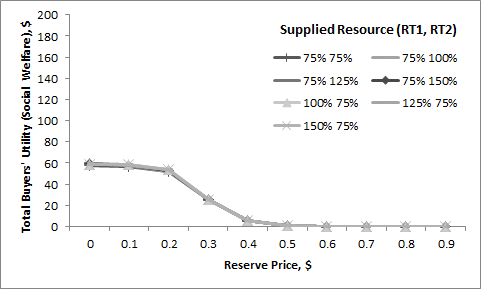} \label{bulUP}
	}
	\subfloat[Buyer's Utility (Social Welfare): OP (Supply $>=100\%$)] {
		\includegraphics[width=0.32\textwidth]{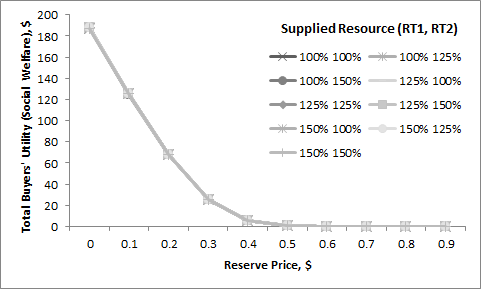} \label{buOP}
	}

	\subfloat[Cost: Supply $RT1=RT2=50\%$, $\varsigma_{run}=25\%$ of $\textit{\={v}}^{mean}$] {
		\includegraphics[width=0.32\textwidth]{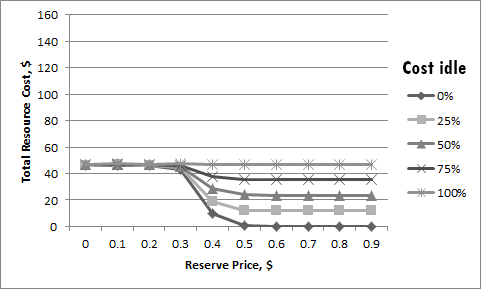} \label{c50}
	}
	\subfloat[Cost: Supply $RT1=RT2=100\%$, $\varsigma_{run}=25\%$ of $\textit{\={v}}^{mean}$] {
		\includegraphics[width=0.32\textwidth]{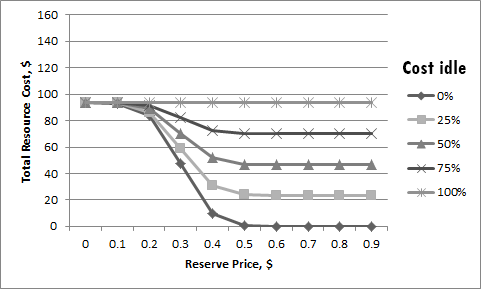} \label{c100}
	}
	\subfloat[Cost: Supply $RT1=RT2=150\%$, $\varsigma_{run}=25\%$ of $\textit{\={v}}^{mean}$] {
		\includegraphics[width=0.32\textwidth]{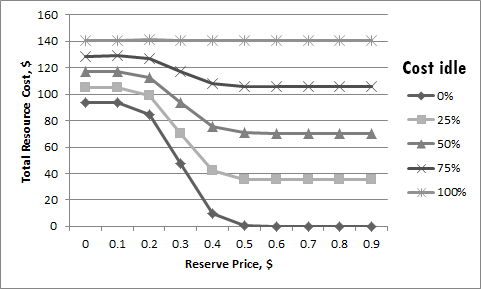} \label{c150}
	}

	\subfloat[Seller's utility: Supply $RT1=RT2=50\%$, $\varsigma_{run}=25\%$ of $\textit{\={v}}^{mean}$] {
		\includegraphics[width=0.32\textwidth]{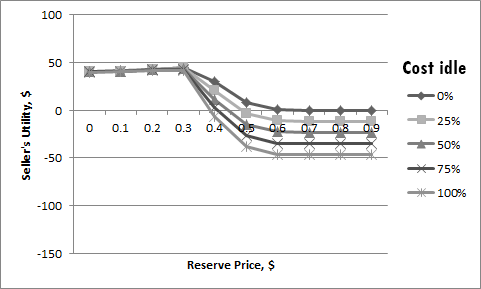} \label{p50}
	}
	\subfloat[Seller's utility: Supply $RT1=RT2=100\%$, $\varsigma_{run}=25\%$ of $\textit{\={v}}^{mean}$] {
		\includegraphics[width=0.32\textwidth]{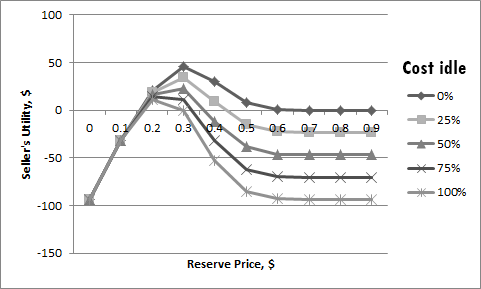} \label{p100}
	}
	\subfloat[Seller's utility: Supply $RT1=RT2=150\%$, $\varsigma_{run}=25\%$ of $\textit{\={v}}^{mean}$] {
		\includegraphics[width=0.32\textwidth]{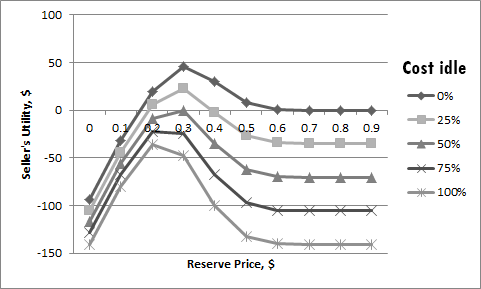} \label{p150}
	}
\end{center}
\vspace{-8px}
\caption{Experiment 2: Greedy-RP Study ($k=2, r_{ij}\in[0,5]$)}
\label{Exp1}
\end{figure*}

Because of the limited space we discuss only representative results. We discuss the case, when 2 resource types are offered in the market. The results are summarized in the Fig. \ref{Exp1}.

\begin{table}[t]
\caption{Experimental Setup}
\centering
\begin{tabular}{c | c}
\hline\hline 
Parameter & Values \\
 \multicolumn{2}{c}{Demand side of the market (Buyers)} \\ [0.5ex]
\hline
Number of bids	& 50	\\
Resource	& 	$r_{ij} \sim \mathcal{N} (2.5, 0.833) : r_{ij} \in [0, 5] $	\\
Valuation	&	$\textit{\={v}}_{j} \sim \mathcal{N} (0.5, 0.166) : \textit{\={v}}_{j} \in [0, 1] $	\\
[1ex]
\hline
 \multicolumn{2}{c}{Supply side of the market (Seller)} \\ [0.5ex]
\hline
Resource Types				& $k=1, 2, 3$	\\
Supply			& 50, 75, 100, 125, 150\% of $r_{ij}$ 	\\ 
$RP$ 					& $o_{i}$ = [0.0, 0.1, ..., 0.9] $\times f_i$ 	\\
\hline
\end{tabular}
\label{ExpSetup2}
\vspace{-5px}
\end{table}

\emph{Supply, Demand and Resource Utilization: }
The Figs. \ref{u1}-\ref{u4} show how the average resource utilization changes with different resource provisioning levels. We can see that the utilization of resources of different type is highly interrelated. An inaccurate cloud provider's estimate for resource supply of one type has a negative impact on the average resource utilization. For example, when the RT1 is supplied in exact quantity (100\%) and the RT2 is under-provisioned (75\%) (Fig. \ref{u1}), the RT2 is fully utilized because of insufficient supply. As the resources in combinatorial bids are complementary, the RT2 becomes a bottleneck and prevents the mechanism from selling the entire capacity of the RT1. Thus, in combinatorial auctions, if at least one type of resource is UP, the entire market becomes UP. We can see that the average utilization is maximized when both resource types are provisioned in equal proportion (e.g. 50\%-50\%, 75\%-75\%) (Figs. \ref{u1}, \ref{u2}).

Across the Figs. \ref{u1} - \ref{u4}, we can see that with increasing $RP$ the average resource utilization decreases. When the $RP$ is high, more bids violate the RPC, which results in reduced resource utilization. However the less resource is provisioned (in equal proportion), the higher the competition between the buyers, and when the $RP$ increases, the resource utilization is higher (Figs. \ref{u3}, \ref{u4}). In other words, when there is not enough resource available, the seller can increase the RPs with the lower negative impact on the resource utilization.

\emph{Revenue and Utilization: }
The total revenue depends on the market state (OP or UP), resource utilization and the $RP$. 
We depict the revenue fluctuation for two UP cases (high UP - Fig. \ref{rhUP} and low UP Fig. \ref{rlUP}), and the OP case (Fig. \ref{rOP}). 
The revenue depends on the amount of sold resource. This amount is bounded either by the market demand in OP case (we cannot sell more than what was requested), or by the least supplied resource in UP case (resource complementarity). In high UP case, compared to low UP case, there is less resource supplied to the market; hence less revenue is generated.

In UP case the revenue generated when $RP=0$ is based on the competition between the buyers. We can see that the seller does not experience significant improvement in her revenue by increasing the $RP$ in UP case. In OP case due to the absence of the competition between the buyers, the $RP$ becomes the price that the bidders pay. As we saw before, higher $RP$ results in reduced resource utilization, which in turn has a negative impact on the revenue (Fig. \ref{rOP}). We can see that the $RP$ is essential for the seller in order to improve the revenue when resource is OP (OP is common for cloud environment).

\emph{Revenue and Buyers' Utility: }
While increasing the $RP$ can have positive and negative impact on the seller's revenue, the buyers' utility (social welfare), which is a total offered discount can only degrade. When the $RP$ is significantly high, buyers' utility drops because of more bids being denied. In fact, the buyers' generated discount and the seller's revenue are directly linked: higher seller's revenues are based on reduced discounts for the buyers (Figs. \ref{rhUP}-\ref{buOP}). The reservation price allows to establish a trade-off between the generated revenue and the total buyers' utility (unlike when $RP=0.0$ \cite{zaman2010})

\emph{Resource Cost: } Any cloud service has a certain incurred cost. In cloud and grid environments it is common to differ the cost of idle resource ($\varsigma_{idle}$) and the cost of the resource which is utilised (i.e. running resource, $\varsigma_{run}$), that tends to be higher because of e.g. increased power consumption \cite{lampe2012}. We calculate the total resource cost, as a sum of costs of running and idle resources. We assume that $\varsigma_{run}=25\% \mbox{ from } \textit{\={v}}^{mean}$(\$0.5) and we vary $\varsigma_{idle}$ as 25, 50, 75, 100\% of $\varsigma_{run}$ (changing $\varsigma_{run}$ will only scale the result). Because of the limited space we depict the graphs for total resource costs when resources of both types are equally UP, EP, and OP (Figs. \ref{c50} - \ref{c150}).

We can see how the total resource cost fluctuates depending on the $RP$, $\varsigma_{run}$ and $\varsigma_{idle}$. With increasing $RP$, the resource utilization drops and more resource becomes idle. As a result, the total cost slopes down as the idle resource is cheaper. The bigger the difference between $\varsigma_{run}$ and $\varsigma_{idle}$, the faster the total cost drops. At the Fig. \ref{c50} (resource is UP) the total cost remains at the same level for the $RP$ from 0.0\$ to 0.3\$ because the utilization is maximized at this interval. At the Fig. \ref{c150} (resource is OP), the resource associated cost is different even when the resource utilization is maximized, because supply $>$ demand (there is always some idle resource).

\emph{Seller's Utility: } We determine seller's utility as the difference between the generated revenue and the total resource associated cost. While the seller's utility mainly depends on the costs of idle and running resource, we depict the example considered for the resource cost 
(Figs. \ref{p50}-\ref{p150}).

We can see that, in all three cases, there is a certain point when seller's utility start to drop. When the resource is UP (Fig. \ref{p50}), the seller's utility remains the same, while the resource utilization is maximized. In case of EP and OP, the seller's utility increases while the revenue increases (Fig. \ref{rOP}). Depending on the values of $\varsigma_{run}$ and $\varsigma_{idle}$, there may be a different strategy for selecting the $RP$ in order to maximize the seller's utility. For example, if  $\varsigma_{idle}$ is very low compared to $\varsigma_{run}$, it may be more interesting to set the $RP$ higher and to sell small amount of resource but at a higher price; the incurred total cost will be small since the idle resource is cheap.

The reservation price allows to establish a trade-off between the seller's and total buyers' utilities. For example, in the case of exact resource provisioning, the $RP=0.3$ results in the total buyers' utility of around \$25 (Fig. \ref{rOP}) and the seller's utility fluctuates between \$0 and \$50 depending on the value of $\varsigma_{idle}$ (Fig. \ref{p100}). In contrast, if there is no RP (or $RP=0.0$), the seller's utility would always be negative which is not acceptable for the seller.

\subsection{Experiment 2 (Allocation Schemes):} In this experiment, we compare three different allocation schemes: the exact optimization mechanism for the Multidimensional Knapsack Problem (MKP) \cite{martello1990} (the recursive formula for dynamic programming implementation was taken from \cite{hans2004}), a modified MKP mechanism with the added RPC (MKP-RP), and a greedy allocation scheme from our proposed mechanism (Greedy-RP). The experimental setting is the same as used in Experiment 1 - Table \ref{ExpSetup2}. We measure the computation time and the social welfare ($\sum_{j:u_j\in W} v_j$), which is a measure of allocative efficiency.

\begin{figure*}[t!]
\setlength{\belowcaptionskip}{-10pt}
\begin{center}
	\subfloat[Computation Time] {
		\includegraphics[width=0.45\textwidth]{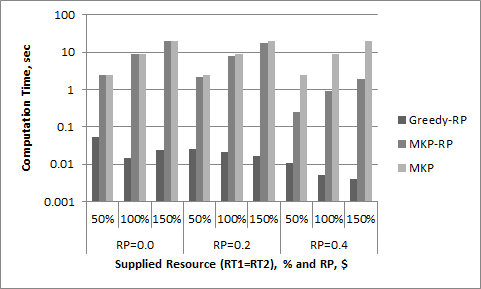} \label{ct}
	}
	\subfloat[Social Welfare: closeness of opt. (Greedy-RP to MKP)] {
		\includegraphics[width=0.45\textwidth]{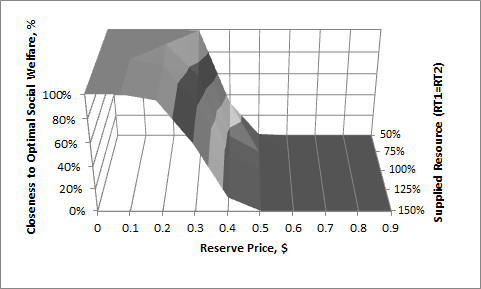} \label{sw}
	}
\end{center}
\vspace{-5px}
\caption{Experiment 2: Allocation Scheme}
\label{Exp1:BuyerUtility}
\end{figure*}

We discuss representative result, when there were two offered resource types ($k=2$). At the Figure \ref{sw}, we demonstrate the closeness of the social welfare, obtained by our proposed mechanism, to the optimal social welfare (MKP). It can be observed that when the $RP$ is increasing, our proposed mechanism produces less efficient results. We noticed that the social welfare is directly linked with the resource utilization: when the utilization is maximized, more bidders obtain the resource; thus, the social welfare is higher. As the MKP does not consider the RPC, the social welfare remains the same (maximized). We can see that when the resource is UP, the optimal allocations are achieved with higher reserve prices, because the resource utilization is close to maximum (Fig. \ref{u3}, case of 50-50\%). If we compare the social welfare of our mechanism with MKP-RP, all the allocations would be close to optimal ($\approx 100\%$).

In terms of computation time, the fastest and the most scalable is the greedy allocation scheme (Fig. \ref{ct}). This result confirms a number of other experiments conducted by other authors \cite{zaman2010}\cite{stober2010}. The MKP-RP computes a little faster compared to the MKP because of the smaller mechanism input (bids that passed RPC only).

\section{Conclusion}
\label{conclusion}
We have designed a truthful market mechanism with the reserve price constraint for trading cloud resources (based on Greedy heuristic). We have investigated the theoretical properties of the mechanism and proved that it is truthful. We have conducted extensive experiments in order to investigate how the amount of supplied and requested resource, and reservation prices influence the cloud resource utilization, generated revenue, resource associated cost, and buyer's utility. Our experiments reveal that using our mechanism in order to maximize the resource utilization, the cloud provider has to provision each resource type in equal proportion to the demand. An inaccurate estimate for the supply of resource of one type has a negative impact on overall resource utilization because of resource complementarity in combinatorial requests. When the resource in the market is under-provisioned, the reserve price, typically, does not achieve significant improvement in the cloud provider's revenue, because the prices are determined based on the competition between the buyers. However, reserve price is essential when the resource is over-provisioned because in such market state a reserve price becomes price-deterministic.
Our experiments show that the reserve price is useful and provides a mechanism to achieve the trade-off between the seller's and the total buyers' utilities. The experiments also show that the allocation scheme in the proposed mechanism yields near optimal allocations and has a low execution time.
In our experiments we have considered the linear bid density. In the future work the impact of various bid densities on the proposed market mechanism's performance will be taken into consideration.

\section*{Acknowledgment}
This work was partially funded by the Smart Services CRC, which is proudly supported by the Australian Federal Government's CRC Grant Program.



\bibliographystyle{IEEEtran}
\bibliography{bib}

\begin{thebibliography}{10}
\providecommand{\url}[1]{#1}
\csname url@samestyle\endcsname
\providecommand{\newblock}{\relax}
\providecommand{\bibinfo}[2]{#2}
\providecommand{\BIBentrySTDinterwordspacing}{\spaceskip=0pt\relax}
\providecommand{\BIBentryALTinterwordstretchfactor}{4}
\providecommand{\BIBentryALTinterwordspacing}{\spaceskip=\fontdimen2\font plus
\BIBentryALTinterwordstretchfactor\fontdimen3\font minus
  \fontdimen4\font\relax}
\providecommand{\BIBforeignlanguage}[2]{{%
\expandafter\ifx\csname l@#1\endcsname\relax
\typeout{** WARNING: IEEEtran.bst: No hyphenation pattern has been}%
\typeout{** loaded for the language `#1'. Using the pattern for}%
\typeout{** the default language instead.}%
\else
\language=\csname l@#1\endcsname
\fi
#2}}
\providecommand{\BIBdecl}{\relax}
\BIBdecl

\bibitem{cloudgrowth}
R.~Gordon, C.~Graham, K.~Hale, J.~Hardcastle, P.~Kjeldsen, G.~Shiffler, and
  E.~Anderson.

\bibitem{chichin2013}
M.~B. Chhetri, S.~Chichin, B.~Q. Vo, and R.~Kowalczyk, ``Smart cloudbench -
  automated performance benchmarking of the cloud,'' in \emph{IEEE Cloud},
  2013.

\bibitem{bakos1998}
Y.~Bakos, ``The emerging role of electronic marketplaces on the internet,''
  \emph{Commun. ACM}, vol.~41, pp. 35--42, 1998.

\bibitem{cortese1998}
A.~Cortese and M.~Stepanek, ``Special report on e-commerce: Goodbye to fixed
  pricing,'' \emph{BusinessWeek}, 1998.

\bibitem{amazonspot}
\BIBentryALTinterwordspacing
``Amazon ec2 spot instances,'' July 2013. [Online]. Available:
  \url{http://aws.amazon.com/ec2/spot-instances/}
\BIBentrySTDinterwordspacing

\bibitem{zaman2010}
S.~Zaman and D.~Grosu, ``Combinatorial auction-based allocation of virtual
  machine instances in clouds,'' in \emph{CloudCom}, 2010, pp. 127--134.

\bibitem{eBayRP}
\BIBentryALTinterwordspacing
``ebay: Selling using reserve price,'' July 2013. [Online]. Available:
  \url{http://pages.ebay.com.au/help/sell/reserve.html}
\BIBentrySTDinterwordspacing

\bibitem{availzones}
\BIBentryALTinterwordspacing
``Amazon ec2 basic infrastructure for windows,'' July 2013. [Online].
  Available:
  \url{http://docs.aws.amazon.com/AWSEC2/latest/WindowsGuide/EC2Win_Infrastructure.html}
\BIBentrySTDinterwordspacing

\bibitem{efficientcombauctions}
T.~Sandholm, ``Algorithm for optimal winner determination in combinatorial
  auctions,'' \emph{Artif. Intell.}, vol. 135, pp. 1--54, 2002.

\bibitem{combauctionapplication}
X.~Zhou, S.~Gandhi, S.~Suri, and H.~Zheng, ``ebay in the sky: strategy-proof
  wireless spectrum auctions,'' in \emph{MOBICOM 2008}.\hskip 1em plus 0.5em
  minus 0.4em\relax ACM, 2008, pp. 2--13.

\bibitem{combauctionapplication2}
Y.~Sheffi, ``Combinatorial auctions in the procurement of transportation
  services,'' \emph{Interfaces}, vol.~34, no.~4, pp. 245--252, 2004.

\bibitem{lehmann2002}
D.~Lehmann, ``Truth revelation in approximately efficient combinatorial
  auctions,'' \emph{Journal of the ACM}, vol.~49, pp. 577--602, 2002.

\bibitem{lena2013}
M.~M. Nejad, L.~Mashayekhy, and D.~Grosu, ``A family of truthful greedy
  mechanisms for dynamic virtual machine provisioning and allocation in
  clouds,'' in \emph{IEEE Cloud}, 2013, pp. 188--195.

\bibitem{stober2010}
J.~St{\"o}{\ss}er, D.~Neumann, and C.~Weinhardt, ``Market-based pricing in
  grids: On strategic manipulation and computational cost,'' \emph{European
  Journal of Operational Research}, vol. 203, pp. 464--475, 2010.

\bibitem{lampe2012}
U.~Lampe, M.~Siebenhaar, A.~Papageorgiou, and R.~S. D.~Schuller, ``Maximizing
  cloud provider profit from equilibrium price auctions,'' in \emph{IEEE
  Cloud}, 2012, pp. 83--90.

\bibitem{fujiwara}
I.~Fujiwara, ``Study on combinatorial auction mechanism for resource allocation
  in cloud computing environment,'' Ph.D. dissertation, The Graduate University
  for Advanced Studies: (SOKENDAI), 2012.

\bibitem{xing-wei2012}
W.~Xingwei, W.~Xue-yi, and H.~Min, ``A resource allocation method based on the
  limited english combinatorial auction under cloud computing environment,'' in
  \emph{FSKD}, 2012, pp. 905--909.

\bibitem{CloudBay}
H.~Zhao, Z.~Yu, S.~Tiwari, X.~Mao, K.~Lee, D.~Wolinsky, X.~Li, and
  R.~Figueiredo, ``Cloudbay: Enabling an online resource market place for open
  clouds,'' in \emph{UCC}, 2012, pp. 135--142.

\bibitem{roovers}
J.~Roovers, ``A market design for iaas cloud resources,'' Master's thesis,
  University of Antwerp, 2011.

\bibitem{amazoninst}
\BIBentryALTinterwordspacing
``Amazon ec2 instance types,'' July 2013. [Online]. Available:
  \url{http://aws.amazon.com/ec2/instance-types/}
\BIBentrySTDinterwordspacing

\bibitem{nisan2007}
N.~Nisan, T.~Roughgarden, E.~Tardos, and V.~Vazirani, \emph{Algorithmic Game
  Theory}.\hskip 1em plus 0.5em minus 0.4em\relax Cambridge University Press,
  2007.

\bibitem{martello1990}
S.~Martello and P.~Toth, \emph{Knapsack problems: algorithms and computer
  implementations}.\hskip 1em plus 0.5em minus 0.4em\relax John Wiley \& Sons,
  Inc., 1990.

\bibitem{hans2004}
H.~Kellerer, U.~Pferschy, and D.~Pisinger, \emph{Knapsack Problems}.\hskip 1em
  plus 0.5em minus 0.4em\relax Springer, 2004, ch.~9.

\end{thebibliography}
%

\end{document}